\documentstyle[12pt,psfig]{article}

\begin{document}
\parindent=0pt

\begin{center}
{\large{\bf A tentative 4$^-$ isomeric state in $^{98}$Sr\\}}
\vspace{3mm}
B. Pfeiffer$^1$, G. Lhersonneau$^{1,a}$, H. Gabelmann$^{1,b}$, 
K.--L. Kratz$^1$, and the ISOLDE-Collaboration$^2$ \\
\vspace{3mm}
$^1$ Institut f\"ur Kernchemie, Universit\"at Mainz\\ 
$^2$ CERN, Geneva, Switzerland\\
\end{center}

In the frame of our systematic investigations of nuclear structure 
properties of 
neutron-rich A$\simeq$100 nuclei at the isotope 
separator ISOLDE at CERN, an isomeric 
level at 1619 keV with the surprisingly 
long $t_{1/2}$ = 85 ns was observed in $^{100}$Sr \cite{pfa,lh01}. 
In analogy to a $K^\pi$=4 rotational band observed 
in the isotone $^{102}$Zr \cite{Durell}, the half-life could 
be interpreted as a $\Delta$K = 4 hindered decay 
from the two quasi-neutron configuration 
($\nu$[411]3/2$\otimes$$\nu$[532]5/2)4$^-$. 
These configurations are supported by quantum Monte-Carlo deformed shell 
calculations, further indicating a strong weakening of the pairing 
residual interaction \cite{cap}.\\

This initiated a reexamination of $\gamma$-$\gamma$-t 
coincidence data on the $\beta$-decay of $^{98}$Rb collected 
at the former OSTIS separator at the ILL (Grenoble) and
during the experiment on $^{100}$Rb decay. 
As a first result, the halflife of the 0$^+_2$ level of $^{98}$Sr
at 215 kev could be remeasured with higher precision \cite{pfb,lh02}.
In addition, an evaluation of the centroid shifts of 
the intense transitions revealed the existence 
of lifetimes related to the transitions 
at 140, 145 and 1693 keV, respectively 
(see Fig.~1 of Ref.~\cite{pfb}). The delay of $t_{1/2}$ = 7 
ns for the 140-1693 keV time distribution 
was attributed to the 1838 keV level (see 
Fig.~2 of Ref.~\cite{pfb}). 
However, $I^\pi$=3$^+$ was assigned to this level after a new analysis 
of the data, leading to a revision of the earlier level scheme of 
$^{98}$Sr \cite{lh02}.
In accordance with quantum Monte-Carlo pairing 
calculations, this isomeric state could 
now be explained as the deformed two quasi-neutron 
configuration ($\nu$[404]9/2$\otimes$$\nu$[411]3/2)3$^+$ 
\cite{lh02}.\\

\begin{figure}
\centerline{\psfig{file=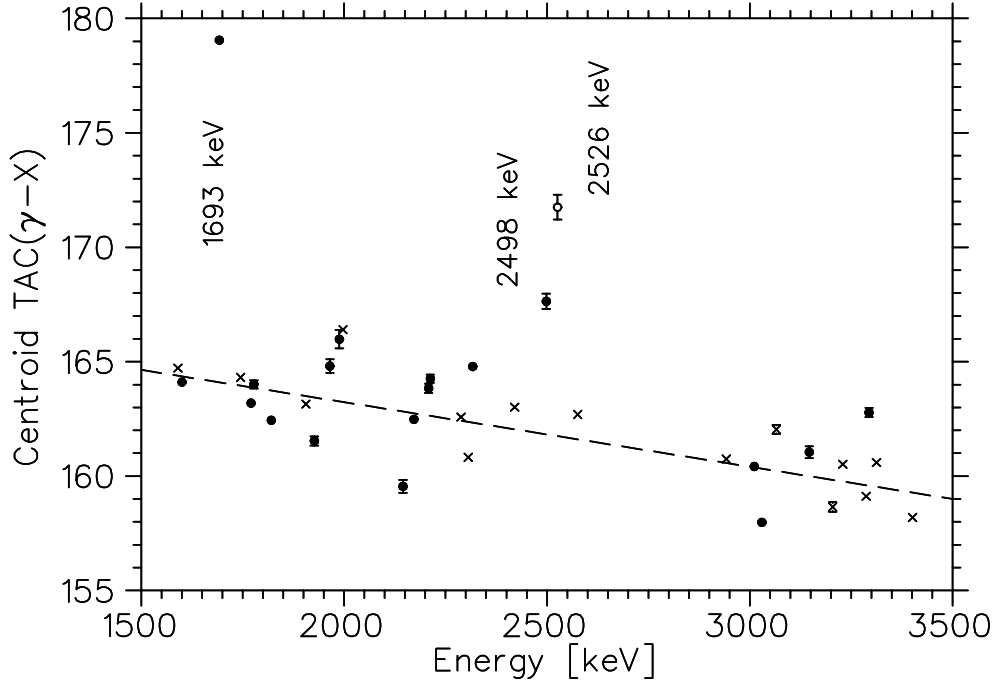,angle=90,width=13.cm}}
\caption{Centroid-shift plot from $\gamma$--$\gamma$--t coincidences in
$^{98}$Sr (filled circles). The scattering of the points gives an indication 
of the systematical errors. The prompt curve (dashed line) is obtained by an
interpolation using transitions in the A=97 and 98 chains (crosses).}
\label{ident}
\end{figure}

In order to search for an analogue to the 4$^-$ 
state at 1619 keV in $^{100}$Sr and 1821 keV in 
$^{102}$Zr, respectively, centroid shifts of higher energy 
transitions are displayed in Fig.~\ref{ident}. This figure 
is an extension to Fig.~3 of Ref.~\cite{lh02} and 
detailed information can be found therein.
Apart from the 1693 kev line, only two transitions show a detectable 
delay: 2498 keV ($t_{1/2}$ = 3$\pm$1 ns) and 2526 
keV ($t_{1/2}$ = 5$\pm$1 ns), respectively. Whereas 
the line at 2498 keV was placed in the decay 
scheme of $^{98}$Sr, the line at 2526 keV 
could (up till now) not be attributed to any 
of the nuclei in the A=97 and 98 decay 
chains following the $\beta$- and $\beta$-delayed 
neutron decay of $^{98}$Rb. K-hindered decays should exist in the 
strongly deformed A=98 Sr and Y isotopes, but feeding from $\beta$-decay 
of the 0$^+$ ground-state of Sr ought not feed
high-spin states in Y. A placement of the 2526 keV line as a
ground-state transition in $^{98}$Sr, possibly fed by a
hinderd low-energy transition, seems a plausible explanation. \\

The 2498 keV line is in coincidence with the 129 and 289 keV
transitions in $^{98}$Sr, 
exclusively deexciting the 
2932 keV level to the 4$^+$ member of the 
ground-state band. 
This single decay branch and the hindrance of the 2498 keV transition 
suggest an analogy with the 1619 keV level in $^{100}$Sr \cite{pfa,lh01}.
The higher excitation energy is in the range expected for such a level 
due to the lowering of the Fermi level for neutrons with respect to 
$^{100}$Sr. Consequently, we propose the 
($\nu$[411]3/2$\otimes$$\nu$[532]5/2)4$^-$ configuration for the 2932 
keV level.
\\

Further studies with new technologies are needed to firmly 
(dis)prove these assumptions.  New measurements ought to be performed 
using the higher beam intensities and better detection systems now
available at CERN-ISOLDE.
With the installation of big Ge-arrays, 
the studies of prompt  deexcitation of fission products have gained new
interest \cite{prog}, offering a complementary approach to decay 
studies. In the particular case discussed here, they could be used to 
search for the band expected to be built on the isomer.

Present address:\\
$^a$ INFN Laboratori Nazionali di Legnaro, Legnaro, Italy\\
$^b$ KSM-Analytik, Mainz
\end{document}